\documentclass[conference]{IEEEtran}
\IEEEoverridecommandlockouts
\usepackage{color}
\definecolor{red}{RGB}{255,0,0}
\usepackage{cite}
\usepackage{amsmath,amssymb,amsfonts}
\usepackage{algorithmic}
\usepackage{graphicx}
\usepackage{textcomp}
\usepackage{stfloats}
\usepackage{xcolor}
\usepackage{stfloats}
\usepackage{booktabs}
\usepackage{cuted}
\def\BibTeX{{\rm B\kern-.05em{\sc i\kern-.025em b}\kern-.08em
    T\kern-.1667em\lower.7ex\hbox{E}\kern-.125emX}}
\begin{document}

\title{Crosstalk-Resilient Beamforming for Movable Antenna Enabled Integrated Sensing and Communication\\
}
\author{Zeyuan Zhang,~Yue Xiu, Zheng Dong, Jiacheng Yin, Maurice J. Khabbaz,~\IEEEmembership{Senior Member,~IEEE}, \\~Chadi Assi,~\IEEEmembership{Fellow,~IEEE},
	~Ning Wei,~\IEEEmembership{Member,~IEEE}\\
	% <-this % stops a space
	\thanks{Z. Zhang, Y. Xiu and N. Wei are with the National Key Laboratory of Wireless Communications, University of Electronic Science and Technology of China, Chengdu 611731, China (e-mail: zzycu@std.uestc.edu.cn;  xiuyue12345678@163.com; wn@uestc.edu.cn).
		
	Z. Dong is with the School of Information Science and Engineering, Shandong University, Qingdao 266237, China (e-mail: zhengdong@sdu.cdu.cn).
	
	J. Yin is with theSchool of Automobile and Transportation, Xihua University, Chengdu 610039, China (yinjiacheng@xhu.edu.cn).
	
	Maurice J. Khabbaz is with the CMPS Department, American University of Beirut, Lebanon (e-mail: mk32l@aub.edu.lb).
		
Chadi Assi is with the Concordia University, Montreal, Quebec, H3G 1M8,Canada (E-mail:chadimassi@gmail.com).
	}
	\thanks{The corresponding author is Ning Wei.}}
\maketitle
\begin{abstract}
This paper investigates a movable antenna (MA) enabled  integrated sensing and communication (ISAC) system under the influence of antenna crosstalk. First, it generalizes the antenna crosstalk model from the conventional fixed-position antenna (FPA) system to the MA scenario. Then, a Cramer-Rao bound (CRB) minimization problem driven by joint beamforming and antenna position  design is presented. Specifically, to address this highly non-convex flexible beamforming problem, we deploy a deep reinforcement learning (DRL) approach to train a flexible beamforming agent. To ensure stability during training, a Twin Delayed Deep Deterministic Policy Gradient (TD3)  algorithm is adopted to balance exploration with reward maximization for efficient and reliable learning. Numerical results demonstrate that the proposed crosstalk-resilient (CR) algorithm enhances the overall ISAC performance compared to other benchmark schemes. 
\end{abstract}
%\begin{IEEEkeywords}
%component, formatting, style, styling, insert
%\end{IEEEkeywords}
%
%\section{Introduction}
%This document is a model and instructions for \LaTeX.
%Please observe the conference page limits. 
\section{Introduction}
Recently, integrated sensing and communication (ISAC) has emerged as a promising technology for future 6G networks \cite{1}. By simultaneously providing communication and sensing functions via a unified waveform and hardware, ISAC successfully reduces the  deployment costs and energy consumption compared to traditional separated radar and communication systems. As a result, ISAC exhibits tremendous potential for development in a variety of prevailing applications such as low-altitude economy, autonomous driving, and smart cities.
%Therefore, ISAC has attracted significant attention of both academia and industrial.
%Through strategic collaboration involving the appropriate reuse of resources and information, ISAC enhances both localization and communication performance [6]. Attributable to these advantages, ISAC have shown great potential in a variety of applications, including vehicle-to-everything (V2X), industrial internet of things (IIoT), and environment monitoring [1]. Moreover, by integrating emerging wireless technologies, such as reconfigurable intelligent surface (RIS) [7], [8], unmanned aerial vehicle (UAV) [9], and non-orthogonal multiple access (NOMA) [10], [11], ISAC has attracted significant attention of both academia and industrial.
 
In conventional Multiple-Input Multiple-Output (MIMO) systems, fixed-position antenna (FPA) is configured and widely employed both for sensing and communication. According to recent works \cite{3,4,5}, by adopting flexible wires and drive components to the traditional FPA systems,  the movable antenna (MA) system offers a promising alternative to improve the performance of FPA-enabled ISAC systems due to the additional degrees of freedom (DoFs) in channel reconstitution. In \cite{4}, the authors  aim to minimize 
the Cramer-Rao Bound (CRB) of angle estimation by jointly optimizing the transmit beamforming
and the position of the MA at the base station (BS). In \cite{5}, the
authors studied the flexible joint beamforming and BS selection scheme with imperfect channel
state information (CSI) in a cooperative ISAC system aided
by movable antennas. The problem is formulated to minimize the transmit power while ensuring that the target
position estimation meets the  CRB requirements and satisfies
the worst-case communication rate constraint. 

However, the performance of multi-antenna systems is significantly limited by the antenna crosstalk \cite{antdpd,75}. For a real wireless transmitter, especially the isolator-free multi-antenna systems like massive MIMO \cite{antdpd}, the transmitted signal at the antenna port is mutually coupled due to the limited isolation between the elements. 
%Moreover, in future transmitter architecture, large-scale multi-antenna systems like massive MIMO comprise up to several hundreds of transmit paths, isolators between PAs and antennas are avoided to reduce system complexity and cost. Therefore, the impact of antenna crosstalk is indispensable for a comprehensive design of MIMO transmitters. 
The antenna crosstalk has been well studied and modeled in the conventional FPA systems.  In \cite{8}, the impact of antenna crosstalk is analyzed and modeled under a uniform linear array (ULA). The authors in \cite{9} show that the crosstalk pattern is related to the physical separation between the antenna elements. However, to the best of the authors’ knowledge, the impact of antenna crosstalk has not been addressed in today's MA enabled ISAC systems, where the antenna movement dynamically affects the crosstalk pattern. 

To fill this gap, this work studies a MA enabled ISAC system under antenna crosstalk. We first  generalize the antenna crosstalk model in conventional FPA to the MA scenario. Then, a CRB minimization problem driven by the flexible beamforming design is formulated. Due to the complexity in the closed-form solution of the problem, an agent-enabled deep reinforcement learning (DRL) approach is adopted for the joint beamforming and antenna position design.  

%The rest of the paper is organized as follows. In Section II, we introduce the MA enabled ISAC system model under antenna crosstalk and formulate the optimization problem. In Section III, we detail the proposed DRL-based transmitter design algorithm. In Section IV, we examine the efficacy of the proposed beamforming scheme via numerical examples. We conclude the paper in Section V.

\section{System Model and Problem Formulation}
\begin{figure}[t]
	\centering
	\includegraphics[width=3.1in]{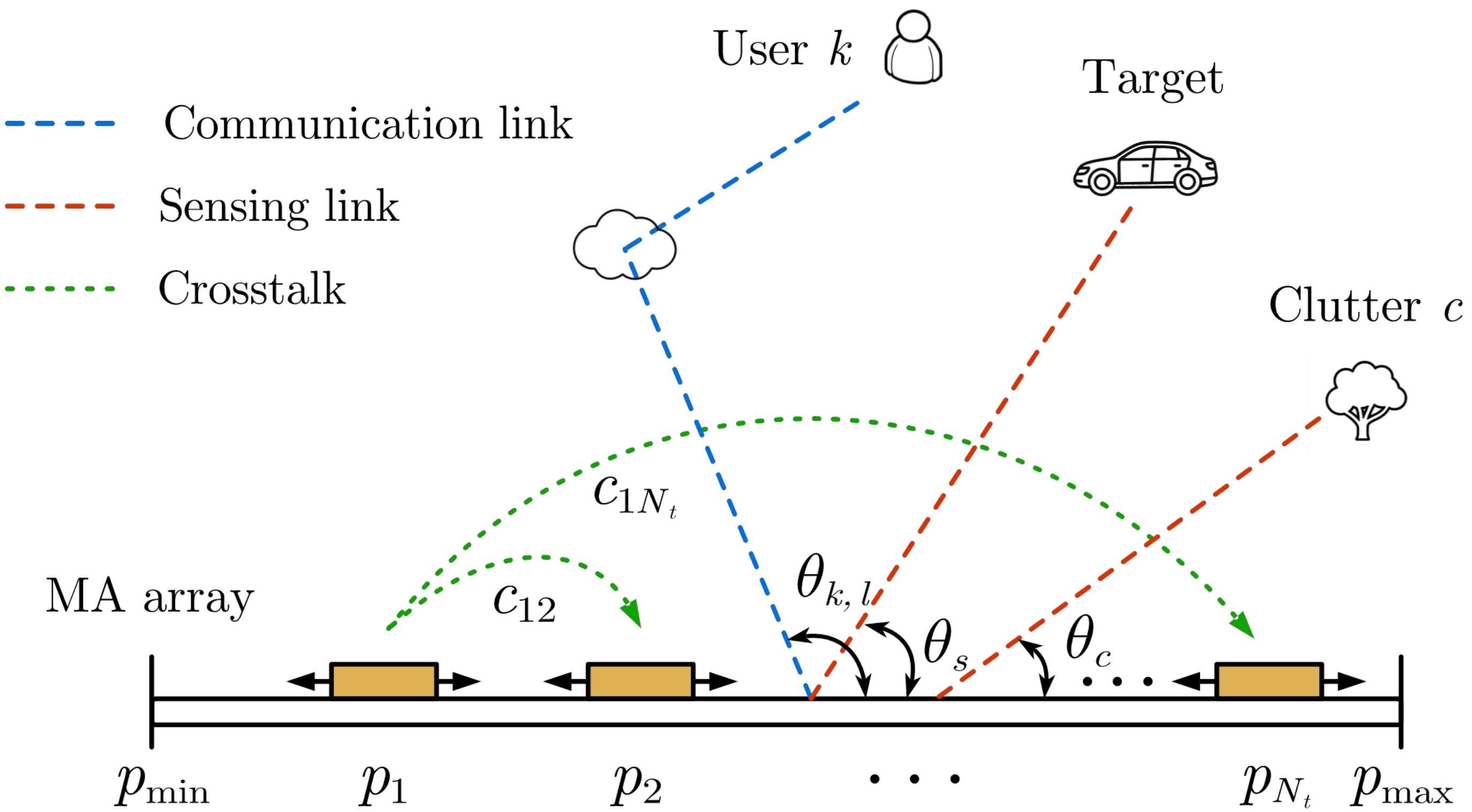}
	\caption{System model of the MA enabled ISAC system under antenna crosstalk.}
	\label{f1}
\end{figure}

As shown in Fig. 1, consider a $ N $-antenna monostatic dual functional radar and communication (DFRC) BS, which simultaneously  serves $ K$  users and senses one target. Each user, as well as the sensing receiver (SR), is equipped with a single antenna. 
\subsection{Channel Model}
According to \cite{3}, the MAs are connected to the radio frequency chains through flexible wires like coaxial cables. This allows the positions of the MAs to be dynamically adjusted using drive components such as stepper motors.  The positions of the transmit MAs are denoted as $ \mathbf{p}=[ p_1,\cdots,p_N]^T  $.  
Under the far-field assumption, the field response vector (FRV) of the $ l $-th propagation path of the $ k $-th user can be expressed as
\begin{equation}\mathbf{a}_{k,l}(\mathbf{p})=\frac{1}{\sqrt{N}}\begin{bmatrix}e^{j\frac{2\pi}{\lambda}p_1 \cos\theta_{k,l}},\cdots,e^{j\frac{2\pi}{\lambda} p_N\cos\theta_{k,l}}\end{bmatrix}^T\in\mathbb{C}^{N\times1},\end{equation}%
where $ \theta_{k,l} $  is the Angle-of-Departure (AoD)  for the $ l $-th path of user $ k $.  Hence, the communication channel between the Tx and user $ k $ can be expressed as
\begin{equation}\mathbf{h}_k(\mathbf{p})=\sqrt{\frac{N}{L_p}}\sum \nolimits _{l=1}^{L_p}\rho_{k,l}\mathbf{a}_{k,l}(\mathbf{p}),\end{equation}%
where $ L_p $ denotes the number of Multi-Path Components (MPC), $ \rho_{k,l} $ is the complex channel gain for the $ l $-th path of user $ k $. In the sequel,  the sensing of a single point target is considered. For proof of concept, a single path model is applied here. With $ \theta_s $ being the target azimuth angle, the FRV from the BS to the sensing target is
\begin{equation}\mathbf{a}_{\kappa}(\mathbf{p})=\frac{1}{\sqrt{N}}\begin{bmatrix}e^{j\frac{2\pi}{\lambda}p_1 \cos\theta_{\kappa}},\cdots,e^{j\frac{2\pi}{\lambda} p_N\cos\theta_{\kappa}}\end{bmatrix}^T\in\mathbb{C}^{N\times1},\end{equation}%
where $ \kappa\in\{s,1,\cdots, C\} $ represents the target or the clutter.
\subsection{Signal Model with Antenna Crosstalk}
Let $ \mathbf{u}[l] \in \mathbb{C}^{N\times1} $ be the precoded DFRC signal at the $ l $-th time slot, which can be written as
\begin{equation}
\mathbf{u}[l]=\mathbf{F}_c\mathbf{s}_c[l]+\mathbf{F}_s\mathbf{s}_s[l],
\end{equation}%
where $ \mathbf{F}_c\in \mathbb{C}^{N\times K}$ and $ \mathbf{F}_s\in \mathbb{C}^{N\times N}$ are the  communication and sensing precoder, respectively. $ \mathbf{s}_c[l]\in \mathbb{C}^{K\times 1}$ and $ \mathbf{s}_s[l]\in \mathbb{C}^{N\times 1}$ are the uncorrelated Wide Sense Stationary (WSS) communication and sensing signals at the $ l $-th slot satisfying $\mathbb{E}\{\mathbf{s}_{c}[l]\mathbf{s}_{c}[l]^{H}\}=\mathbf{I}_{K}$ and $\mathbb{E}\{\mathbf{s}_{s}[l]\mathbf{s}_{s}[l]^{H}\}=\mathbf{I}_{N}$. For simplicity, we define $\mathbf{F}\triangleq[\mathbf{F}_{\mathrm{c}}\, \mathbf{F}_{\mathrm{s}}]$ and $\mathbf{s}[l]\triangleq[\mathbf{s}_\mathrm{c}[l]^T\,\mathbf{s}_\mathrm{s}[l]^T]^T$. It then holds that
	\begin{equation}
		\mathbf{u}[l]=	\mathbf{F}	\mathbf{s}[l].
	\end{equation}
As shown in Fig. 1, under the effect of antenna crosstalk, the actual signal at the output of antenna $ n $ is a mixture of signals from all branches, which can be expressed as in \cite{co}
	\begin{equation}
		x_n[l]=\sum \nolimits_{m=1}^N c_{mn}u_m[l],
	\end{equation}
where $ x_n[l] $ and $ u_m[l] $ are the $ n $-th and $ m $-th components of $ \mathbf{x}[l] $ and $ \mathbf{u}[l] $, respectively. $ c_{mn} $ is the coupling coefficient from antenna $ m $ to antenna $ n $. According to \cite{8}, the coupling coefficients can be described by a linear phase power series, that is, $ c_{mn} $ can be well modeled as
	\begin{equation}c_{mn}=\eta\cdot d_{mn}^{-\iota}\cdot e^{-j(\nu d_{mn}+\xi)},\end{equation}%
where $ \eta$, $ \iota$, $ \nu $ and $\xi $ are model parameters, $ d_{mn} = \left| p_m-p_n\right|, m,n\in\{1,\ldots,N\}$ is the separation between the $ m $-th and the $ n $-th antenna. 
Accordingly, in the context of the MA array, the  transmitted signal matrix is given by
\begin{equation}\mathbf{X}=\mathbf{C}(\mathbf{p})\mathbf{F}\mathbf{S},\end{equation}%
where $ \mathbf{X} =[\mathbf{x}[1],\cdots,\mathbf{x}[L]]$, $ \mathbf{S} =[\mathbf{s}[1],\cdots,\mathbf{s}[L]]$, $ \mathbb{E}\{\mathbf{SS}^H\} =L \cdot \mathbf{I}_{N+K}$. $ \mathbf{C}(\mathbf{p})\in\mathbb{C}^{N\times N} $ is the coupling matrix with $ [\mathbf{C}(\mathbf{p})]_{mn}=c_{mn} $.
After spatial  combining through the channel,  the received signal for user $k  $ at time slot $ l $ can be written as
\begin{equation}
	y_{k}[l]={\mathbf{h}^H_k(\mathbf{p}) \mathbf{C}(\mathbf{p})\mathbf{f}_k s_k[l]}+{\mathbf{h}^H_k(\mathbf{p}) \mathbf{C}(\mathbf{p})\sum_{k^{\prime}\neq k}\mathbf{f}_{k^{\prime}}s_{k^{\prime}}[l]}+{n_k[l]},
\end{equation}%
where $ s_k[l] $ is the $ k $-th component of $ \mathbf{s}[l] $, $n_{k}[l]\sim\mathcal{CN}(0,\sigma^{2}_{k})$ is the Additive White Gaussian Noise (AWGN) at user $ k $. 
%\begin{equation}y_k=\mathbf{h}_k^H\phi(\mathbf{x})+n_k,\end{equation}
Provided that the frame length $ L $ is sufficiently large, the Signal-to-Interference-plus-Noise Ratio (SINR) of user $ k $ is given by
	\begin{equation}\mathrm{\gamma}_k=\frac{|\mathbf{h}^H_k(\mathbf{p}) \mathbf{C}(\mathbf{p})\mathbf{f}_k|^2}{\sum_{k^{'} \neq k}|\mathbf{h}^H_k(\mathbf{p}) \mathbf{C}(\mathbf{p})\mathbf{f}_{k^{\prime}}|^2+\sigma_k^2}.\end{equation}%
The echo signal received by the SR can be expressed as
\begin{equation}
		\begin{aligned}
	y_s[l]={\alpha_s\mathbf{a}_s^H(\mathbf{p})\mathbf{C}(\mathbf{p})\mathbf{F} \mathbf{s}[l]}+{\sum \nolimits_c\alpha_c\mathbf{a}_c^H(\mathbf{p})\mathbf{C}(\mathbf{p})\mathbf{F} \mathbf{s}[l]}+{n_s[l]},
\end{aligned}\end{equation}%
where $ \alpha_s $ and  $ \alpha_c $ are  the complex coefficients of the Radar Cross Section (RCS) of the target and the clutter $ c $, respectively.  $n_s[l]\sim\mathcal{CN}(0,\sigma_s^2)$ is the AWGN of the sensing link. 
\subsection{Performance Metric and Problem Formulation}
Consider the CRB for estimating $ \theta_s $ as the metric to evaluate the sensing performance, it is given by
 \begin{eqnarray}
 	\begin{aligned}
 		\mathrm{CRB}_{\theta_s} =\frac{{\sigma_n^2}}{
 			{2L|\alpha_s|^2}\left( 
 			 \dot{\mathbf{g}}^H_s(\mathbf{p}) \mathbf{F}\mathbf{F}^H\dot{\mathbf{g}}_s (\mathbf{p}) 
 			- \frac{ \left|   \mathbf{g}^H_s(\mathbf{p})  \mathbf{F} \mathbf{F}^H \dot{\mathbf{g}}_s(\mathbf{p})   \right|^2}
 			{{\mathbf{g}}^H_s(\mathbf{p}) \mathbf{F}\mathbf{F}^H{\mathbf{g}}_s(\mathbf{p}) }
 			\right)
 		},
 	\end{aligned}\label{crb}
 \end{eqnarray}%
where  $\mathbf{g}_{s}(\mathbf{p})=\mathbf{C}^H(\mathbf{p})\mathbf{a}_s(\mathbf{p})$ denotes the effective crosstalk-involved channel. 
The derivation of the CRB under antenna crosstalk is given in the \textbf{Appendix}. This paper  aims at minimizing the CRB for estimating $ \theta_s $ by jointly designing the transmit precoding $ \mathbf{F} $ and antenna position $ \mathbf{p} $ while assuring the QoS requirements of communication users. The optimization problem can be formulated as
%\begin{alignat}&\min_{\mathbf{F}_{\mathrm{A}},\mathbf{F}_{\mathrm{D}},\mathbf{p}}\mathrm{CRB}(\theta_s)\\&\mathrm{s.t.}\,\,\gamma_{k}\geq\Gamma_{k},\forall k,\\&\mathbf{p}\in\mathcal{C},\\&\mathbf{F}_{\mathrm{A}}\in\mathcal{A},\\&\mathbb{E}\left[\left\|\phi(\mathbf{F}\mathbf{s})\right\|^2\right]=P.\end{alignat}
\begin{subequations}
	\begin{align}
		\min_{\mathbf{F},\mathbf{p}}&~ \mathrm{CRB}_{\theta_s}\label{14}\\
		\mbox{s.t.}~
		&~|p_m-p_n|\geq D_0,\forall m,n\in\{1,\ldots,N\},m\neq n,\label{14a}\\
		&~p_{\mathrm{min}}\leq p_m\leq p_{\mathrm{max}}, \forall m\in\{1,\ldots,N\},\label{14b}\\
		&~\gamma_k\geq\Gamma_k, \forall k \in \{1,\cdots,K\},&\label{14c}\\
		&~\mathrm{Tr}(\mathbf{F}\mathbf{F}^H)\leq P_{\rm sum},&\label{14d}
	\end{align}\label{143}%
\end{subequations}%
where constraint (\ref{14a}) ensures the  sufficient separation $ D_0 $ is fulfilled between the antenna elements; (\ref{14b}) corresponds to the feasible moving region of the MAs; Constraint (\ref{14c}) guarantees the sufficient SINR of communication users, while constraint (\ref{14d}) limits the overall transmit power.
\section{Proposed Crosstalk-Resilient Beamforming with Movable Antenna}
To address the complex optimization problem in \eqref{143} for MA
enabled ISAC system under antenna crosstalk,  a DRL-based algorithm for flexible beamforming is proposed based on the Twin Delayed Deep Deterministic Policy Gradient (TD3)  framework \cite{td3}. The DRL problems are often modeled as Markov Decision Processes (MDPs), defined
by the tuple $ (S,A,P,R,\varrho) $,  where $ S $ and $ A $ are the state and
action spaces, $ P $ is the transition function, $ R $ is the reward, and $ \varrho $ is the discount factor. The agent follows a policy $\pi  $ to maximize the cumulative reward across multiple training episodes.
%$ G(t)=\sum_{i=0}^{T-t-1}\gamma^{i}r_{t+i+1} $
%across each episode. The goal is to find an optimal
%policy  $\pi^*=\arg\max_\pi\mathbb{E}_\pi\left[G(t)\right]$.
\subsection{MDP Formulation}
\subsubsection{State space}
The state space provides the flexible beamforming agent with essential information for the policy selection in a concise form
\begin{equation}
	\mathbf{s}_t = \{\Re\left\lbrace  \boldsymbol{\rho} \right\rbrace, \Im\left\lbrace  \boldsymbol{\rho} \right\rbrace ,\boldsymbol{\theta} , \mathbf{a}_{t-1}\},
\end{equation}%
where $ \boldsymbol{\rho} $ and $ \boldsymbol{\theta} $ refer to the complex channel gain coefficients and the azimuth angles, specifically given by $ \boldsymbol{\rho} = [\boldsymbol{\rho}^T_1, \boldsymbol{\rho}^T_2, \cdots, \boldsymbol{\rho}^T_K]^T $, $ \boldsymbol{\rho}_k = [{\rho}_{k,1}, {\rho}_{k,2}, \cdots,{\rho}_{k,L_p}]^T $, $ \boldsymbol{\theta} = [\boldsymbol{\theta}^T_1, \boldsymbol{\theta}^T_2, \cdots, \boldsymbol{\theta}^T_K]^T $, $ \boldsymbol{\theta}_k = [{\theta}_{k,1}, {\theta}_{k,2}, \cdots,{\theta}_{k,L_p}]^T $.
Since the neural networks can only handle real numbers, the channel characteristics are represented as separate real and imaginary parts. 
\subsubsection{Action space}
 The joint beamforming and  antenna positioning agent adjusts the action defined by
 \begin{equation}
 	\mathbf{a}_t = \{\mathbf{a}^{\mathbf{F}}_t ,\mathbf{a}^{\mathbf{p}}_t\},
 \end{equation}%
 where $ \mathbf{a}^{\mathbf{F}}_t = \Re\left\lbrace \mathrm{vec}\left( \mathbf{F}_t\right) \right\rbrace, \Im\left\lbrace \mathrm{vec}\left( \mathbf{F}_t\right) \right\rbrace $ and $ \mathbf{a}^{\mathbf{p}}_t = \mathbf{p}_t $ are the beamforming and  antenna  action at time $ t $, respectively.
\subsubsection{Reward}
The reward function provides direct feedback
from the environment to guide the policy selection of the joint agent. It is structured as
\begin{equation}
R_t=  \dot{\mathbf{g}}^H_s(\mathbf{p}_t)\mathbf{F}_t\mathbf{F}^H_t\dot{\mathbf{g}}_s(\mathbf{p}_t) -\frac{ \left|   \mathbf{g}^H_s(\mathbf{p}_t) \mathbf{F}_t \mathbf{F}^H_t \dot{\mathbf{g}}_s(\mathbf{p}_t)  \right|^2}
{{\mathbf{g}}^H_s(\mathbf{p}_t)\mathbf{F}_t\mathbf{F}^H_t{\mathbf{g}}_s(\mathbf{p}_t) }+ c_t^{\mathrm{SINR}},
\end{equation}%
where 
\begin{equation}c_t^{\mathrm{SINR}} = \upsilon\cdot\sum_{i=1}^K\min\left( 0,\left(\gamma_i(t)-\Gamma_i\right)\right) \end{equation}%
is the penalty term corresponding to the constraint \eqref{14c}, $ \upsilon $ is the penalty factor, $ \gamma_i(t) $ is the SINR of user $ i $ calculated by $  \mathbf{F}_t $ and $ \mathbf{p}_t $. 
As far as this work is concerned, other constraint conditions are separately combined with their own agent. For instance, the power constraint \eqref{14d} can be managed through a custom
activation layer. 
%transforming the input as
%\begin{equation}\mathbf{f}_k=P_{sum}\cdot\left(\frac{e^{\boldsymbol{\mu}^\top\boldsymbol{a}_k}}{\sum_{i=1}^K\boldsymbol{\mu}^\top\boldsymbol{a}_i}\right)\times e^{\mathrm{j}\cdot2\pi\boldsymbol{\mu}^\top\boldsymbol{a}_k}.\end{equation}
This corresponds to $ N\times K $ unit modulus phase achieved by $ \mathtt{tanh}  $ and $ K $ power control coefficients achieved by $ \mathtt{softmax} $.  For  antenna position optimization, a group of  displacement variables $\{\Delta_{n}\}_{n=1}^{{N}}$ with recursive position relation is introduced to simplify the network design. For the $ n $-th antenna, the position is given by  
\begin{equation}
	\left\{
	\begin{array}{lr}
		p_1 = p_{min} + \Delta_1, &  \\
		p_n=p_1+(n-1)D_0+\sum_{k=1}^n\Delta_k,&(2\leq n\leq N),\\ 
	\end{array}
	\right.
\end{equation}
By introducing the displacement variables, the antenna position constraint \eqref{14a} and \eqref{14b} can be equivalently written as $\{\Delta_{n}\}_{n=1}^{{N}}\geq0$ and $\sum_{n=1}^{N}\Delta_{n}\leq\Delta_{max}$ with $ \Delta_{max} = p_{max}-(N-1)D_0 $, which can be implemented through $ \mathtt{sigmoid} $ and $ \mathtt{softmax} $ functions.
\subsection{TD3 Training }
For the agent training, a TD3 based framework is employed. TD3 is an off-policy actor–critic method, where the actor outputs deterministic actions and two critic networks estimate their Q-values. The learning process alternates between interaction with the environment and parameter updates using mini-batches sampled from a replay buffer $ \mathcal{D} $. During training, the agent updates the two critic networks $Q_1  $ and $ Q_2 $ by minimizing the mean squared Bellman error, which is expressed as
\begin{equation}\mathcal{L}_{Q_i}=\mathbb{E}_{(s,a,r,s^{\prime})\sim\mathcal{D}}\left[(Q_i(s,a)-y)^2\right],\quad i=1,2,\end{equation}
where $
y $ is the target Q-value calculated by $ r + \gamma \min_{j=1,2} Q'_j(s', \tilde{a}^{\prime})  $, $ \tilde{a}^{\prime} $ is the smoothed target action obtained from the target actor network. To further improve the stability, target policy smoothing noise is added to the action, which is given by
\begin{equation}
	\tilde{a}^{\prime} =  \pi^{\prime}(s^{\prime})+\varepsilon,\end{equation}
where $ \varepsilon $ is the clipped Gaussian noise with
$ \varepsilon \sim \operatorname{clip}\!\left(\mathcal{N}(0,\sigma_{cn}^{2}),\, -c,\, c\right) $. 
The actor policy 
$ \pi_\phi(s) $ is optimized to maximize the Q-value predicted by the first critic network
\begin{equation}\max_\phi\mathcal{J}(\phi)=\mathbb{E}_{s\sim\mathcal{D}}[Q_{\mu_1}(s,\pi_\phi(s))].\end{equation}%
The policy actor network and its target counterpart are updated at a reduced frequency compared to the critic networks, with a common ratio of 0.5. For both the target actor and critic networks, the  parameter is updated using soft updates 
\begin{equation}\begin{array}{l}\mu_i^{\prime}\leftarrow\tau\mu_i+(1-\tau)\mu_i^{\prime},\quad \phi^{\prime}\leftarrow\tau\phi+(1-\tau)\phi^{\prime}, \quad i=1,2,\end{array}\end{equation}%
where $ \tau $ is the soft update coefficient. 
In the proposed TD3 algorithm, in order to unify the noise scale of different optimization variables and avoid the gradient vanishing due to extra clipping and projection,  the Ornstein-Uhlenbeck (OU) noise is applied before the logits are mapped into concrete actions. Furthermore, to ensure the sufficient exploration in the early training phase while  reducing the stochastic perturbations when converge to the optimal deterministic policy, the OU noise  power is configured to decay in an exponential manner, as such, it satisfies
\begin{equation}
	\sigma_{ou}(\zeta) = \sigma_{ou,0} e^{-\varpi \zeta} + \sigma_{ou,\min},
	\end{equation}%
where $  \zeta $ is the episode index, $ \varpi $ is the noise decay factor, $ \sigma_{ou,0} $ and $ \sigma_{ou,\min} $ are the initial and minimum noise power, respectively. 
\section{Numerical Result}
\begin{table}[t]
	\centering
	\caption{Simulation Parameters}
	\begin{tabular}{@{}lcl@{}}
		\toprule
		\textbf{Parameter} & \textbf{Value} & \textbf{Description} \\
		\midrule
		$K$ & $4$ & Default number of users \\
		$L_p$ & $3$ & Default MPC  number\\
		$f_{c}$ & $30$ & Carrier frequency [GHz] \\
		$p_{min}$ & $0$ & Lower boundary of the movable region [m]\\
		$p_{max}$ & $0.15$ & Upper boundary of the movable region [m]\\
		$L$ & $128$ & Estimation samples \\
		$\alpha_{s}$ & $0.4$ & RCS of the target \\
		$\sigma_{cn}^{2}$ & $0.15$ & Clipped Gaussian noise variance \\
		$c$ & $0.01$ & Clipped Gaussian noise range \\
		$\varrho$ & $0.98$ & Discount factor \\
		$\tau$ & $0.003$ & Soft update coefficient \\
		$ \sigma_{ou,0} $& $0.1$ & Initial OU noise power \\
		$ \sigma_{ou,\min} $& $0.005$ & Minimum OU noise power \\
		$ \varpi $ & $0.006$ & Noise decay rate \\
		$ \upsilon $ & $0.1$ & Penalty factor \\
		\bottomrule
	\end{tabular}
\end{table}
\begin{figure}[t]
	\centering
	\includegraphics[height= 1.5in, width=3.2in]{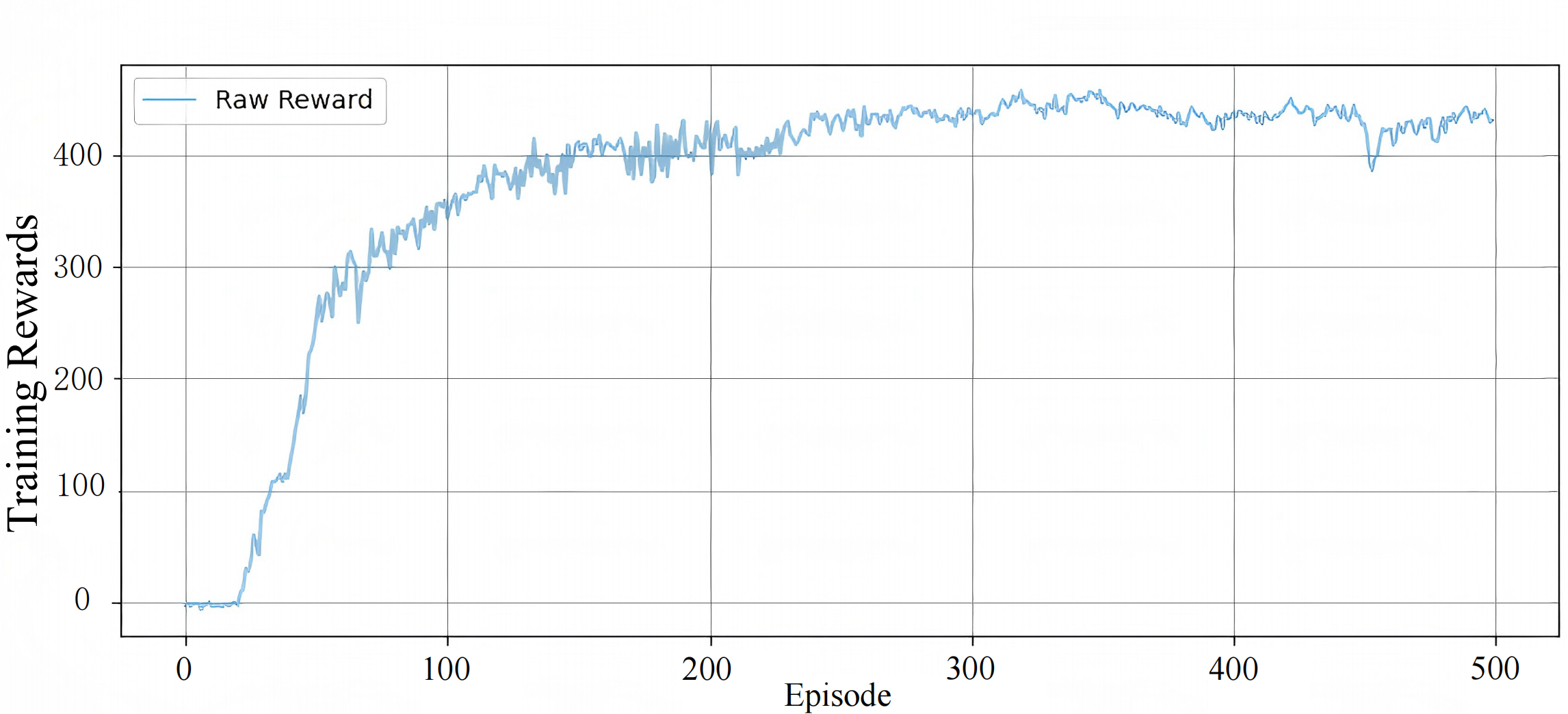}
	\caption{ Training convergence of the proposed TD3 algorithm.}
	\label{f1}
\end{figure}
%\begin{figure}[t]
%	\centering
%	\includegraphics[width=3.1in]{1}
%	\caption{System model of the hybrid beamforming ISAC with nonlinear amplification.}
%	\label{f1}
%\end{figure}
%\begin{figure}[t]
%	\centering
%	\includegraphics[width=3.1in]{1}
%	\caption{System model of the hybrid beamforming ISAC with nonlinear amplification.}
%	\label{f1}
%\end{figure}
This section evaluates a 16-MA ISAC system operating under antenna crosstalk. For $ K $ users, the locations are generated through random channel realization with $ \rho_{k,l} \sim \mathcal{CN}(0,1)$ and $ \theta_{k,l} \sim \mathcal{U}(0,\pi)$. Users and the target are considered to have the same noise level and weighting factors, (i.e., $ \sigma_k^2=\sigma_s^2=N_0 $). The transmit power $  P_{sum} $ is configured at 10 dBm, and the Signal-to-Noise Ratio (SNR) is given by $ P_{sum}/N_0 $. Throughout the experiments, a single target is considered to be located at $ \theta_s =60^{\circ}$. 
In each training episode, 100 steps are employed to engage the interaction with the environment. For the antenna crosstalk, we employ the closed-form model deduced from \cite{8}, where the model parameters are given by $ \eta = 3.5\times10^{-5} $, $ \iota =  1.9 $, $ \nu = 600.4 $, $ \xi = 252.8$. Other simulation parameters are listed in Table I.

\begin{figure}[t]
	\centering
	\includegraphics[width=2.5in]{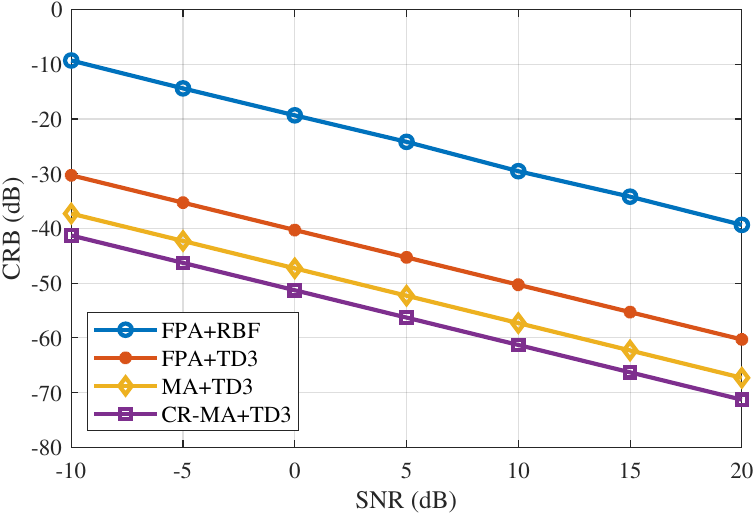}
	\caption{Average CRB performance versus SNR.}
	\label{f1}
\end{figure}
\begin{figure}[t]
	\centering
	\includegraphics[width=2.5in]{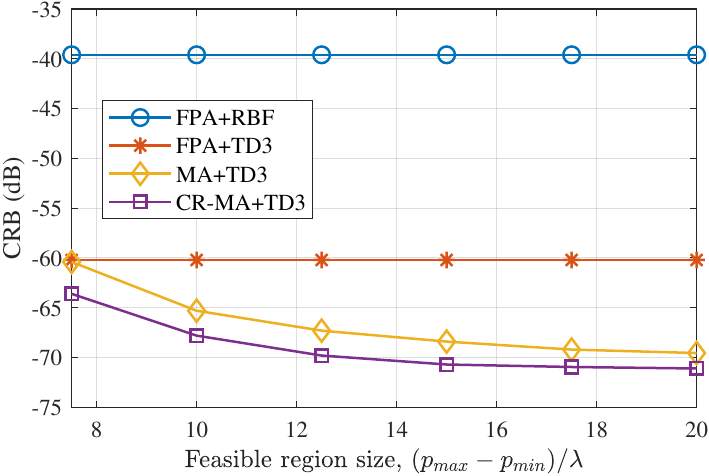}
	\caption{Average CRB performance versus the different region of the array.}
	\label{f1}
\end{figure}

Fig. 2 illustrates the training convergence of the proposed TD3 algorithm. As can be observed, thanks to the twin critic network and delayed update of action selection, the proposed TD3 learning shows stable convergence around 200 episodes and vibrates around 430 after that. Specifically, the first 30 episodes are reserved as a warm-up procedure to avoid the agent from blindly pursuing the gradient direction.  The result validates TD3’s capability to handle the joint variable optimization, offering a scalable solution for MA enabled ISAC.

Fig. 3 illustrates the average CRB achieved by four strategies under varying SNR. The FPA+RBF corresponds to the scheme where random beamforming (RBF) is applied to a fixed ULA with half-wavelength spacing. The FPA+TD3 corresponds to the scheme where the agent optimizes only the beamforming matrix $ \mathbf{F} $ and the array is fixed as a ULA with  half-wavelength spacing. MA+TD3 utilizes the proposed TD3 training  without considering the antenna crosstalk, while the CR-MA+TD3 corresponds the proposed crosstalk-resilient MA scheme using TD3. Observe that the CRB in dB of all schemes
decreases monotonically in an approximately linear manner,  this is because a higher
SNR results in a better signal quality in target estimation. It is also noted that the proposed CR-MA+TD3 outperforms all alternatives, since it considers the actual antenna crosstalk into consideration and performs a successful joint  optimization through the TD3 agent.
 
 Fig. 4 describes the average CRB variation versus the size of MA moving region. The proposed CR-MA scheme  exhibits a substantial performance gain, with the CRB rising from -63.8 dB at a 7.5$ \lambda $ configuration to
 over -70.9 dB when expanded to 20$ \lambda $.   In Fig. 4, it is shown that the CRB of MA enabled angle estimation slowly
 decreases as the moving region increases, this improvement
 stems from increased system DoFs, which enables optimized
 beamforming region.
 
\section{Conclusion}
This paper addresses the crosstalk-resilient flexible beamforming problem in MA enabled ISAC system. For this system, a crosstalk-resilient flexible beamforming algorithm is proposed to minimize the CRB of angle estimation while meeting the
antenna position, user QoS, and transmit power constraints. To solve the highly non-convex 
optimization problem,  a TD3-based DRL approach  is propsoed, where a flexible beamforming agent is trained to learn the policy of joint optimization of both antenna positioning and beamforming. According to the  numerical results, by introducing the MA, the performance of the ISAC system under antenna crosstalk can be improved significantly.

\appendix
%Given the clutters’ directions, the parameters to be estimated is denoted by $ \boldsymbol{\psi}\triangleq[\theta_{s},\boldsymbol{\alpha}_s^T]^{T} $ with $ \boldsymbol{\alpha}_s=[\Re\{\alpha_s\},\Im\{\alpha_s\}]^{T} $. Let $ \mathbf{y}_s=[{y}_s[1],\cdots,{y}_s[L]]^T $ be the complex observation of the sensing signal. The covariance matrix of the clutter echo is $ 	\mathbb{E} \left\{ |\alpha_c|^2 \mathbf{S}^H \mathbf{F}^H \mathbf{a}_c \mathbf{a}_c^H \mathbf{F} \mathbf{S} \right\} 
%=   {\sigma}^2_c \cdot \mathbf{I}_L $, where $ {\sigma}^2_c =|\alpha_c|^2 \mathbf{a}_c^H \mathbf{F} \mathbf{F}^H \mathbf{a}_c $. Thus, we have $ \mathbf{y}_s\sim \mathcal{CN}(\bar{\mathbf{y}}_s,\mathbf{R}_n)$ with $ \bar{\mathbf{y}}_s={\alpha^*_s\mathbf{S}^H\mathbf{F}^H\mathbf{g}_s  }$ and $ \mathbf{R}_n= {\sigma}^2_n \cdot \mathbf{I}_L$.  Where  $\mathbf{g}_{s}(\mathbf{p})=\mathbf{C}^H(\mathbf{p})\mathbf{a}_s(\mathbf{p})$ denotes the effective crosstalk-involved channel, $ {\sigma}^2_n={\sigma}^2_s +{\sigma}^2_c $ is the effective noise variance. We have
Given the clutters’ directions, the parameters to be estimated is denoted by $ \boldsymbol{\psi}\triangleq[\theta_{s},\boldsymbol{\alpha}_s^T]^{T} $ with $ \boldsymbol{\alpha}_s=[\Re\{\alpha_s\},\Im\{\alpha_s\}]^{T} $. Let $ \mathbf{y}_s=[{y}_s[1],\cdots,{y}_s[L]]^T $ be the complex observation of the sensing signal. According to \cite{13}, the clutter echo can be viewed as zero-mean Gaussian noise with power $ {\sigma}^2_c $.  Thus, we have $ \mathbf{y}_s\sim \mathcal{CN}(\bar{\mathbf{y}}_s,\mathbf{R}_n)$ with $ \bar{\mathbf{y}}_s={\alpha^*_s\mathbf{S}^H\mathbf{F}^H\mathbf{g}_s(\mathbf{p})  }$ and $ \mathbf{R}_n= {\sigma}^2_n \cdot \mathbf{I}_L$. Let  $\mathbf{g}_{s}(\mathbf{p})=\mathbf{C}^H(\mathbf{p})\mathbf{a}_s(\mathbf{p})$ denote the effective crosstalk-involved channel, and $ {\sigma}^2_n={\sigma}^2_s +{\sigma}^2_c $ is the effective noise variance. It follows that
\begin{subequations}
	\begin{align}
		\frac{\partial\bar{\mathbf{y}}_s}{\partial\theta_s}&=\alpha^*_s\mathbf{S}^H\mathbf{F}^H\dot{\mathbf{g}}_s(\mathbf{p})\in\mathbb{C}^{L\times1},\\	\frac{\partial\bar{\mathbf{y}}_s}{\partial\boldsymbol{\alpha}_s}&=\boldsymbol{\epsilon}\otimes (\mathbf{S}^H\mathbf{F}^H\mathbf{g}_s(\mathbf{p}))\in\mathbb{C}^{L\times2},
	\end{align}
\end{subequations}%
where $ \boldsymbol{\epsilon}=[1,j] $, 
%$ \dot{\mathbf{g}}^H_s $ denotes the partial derivative of $ \mathbf{g}^H_s $ w.r.t. $ \theta_s $, which is derived as 
$ \dot{\mathbf{g}}_s(\mathbf{p}) $ denotes the partial derivative of $ \mathbf{g}_s $ w.r.t. $ \theta_s $, which is derived as 
%\begin{equation}
% \dot{\mathbf{g}}^H_s= \frac{\partial\mathbf{g}^H_s}{\partial\theta_s}=  j \frac{2\pi}{\lambda} \sin \theta_s \cdot \mathbf{a}^H_s(\mathbf{p}) \cdot \operatorname{diag}(\mathbf{p}) \cdot\mathbf{C}(\mathbf{p}).\end{equation}% 
\begin{equation}
\dot{\mathbf{g}}_s(\mathbf{p})= \frac{\partial\mathbf{g}_s(\mathbf{p})}{\partial\theta_s}=-  j \frac{2\pi}{\lambda} \sin \theta_s \cdot\mathbf{C}^H(\mathbf{p})\cdot\operatorname{diag}(\mathbf{p}) \cdot \mathbf{a}_s(\mathbf{p}) .\end{equation}% 
Consequently, the 
Fisher information matrix $ \mathbf{M}\in\mathbb{C}^{3\times3} $ is
\begin{equation}\mathbf{M}=\begin{bmatrix}\mathbf{M}_{\theta_s\theta_s}&\mathbf{M}_{\theta_s\boldsymbol{\alpha}_s}\\\mathbf{M}_{\boldsymbol{\alpha}_s\theta_s}&\mathbf{M}_{\boldsymbol{\alpha}_s\boldsymbol{\alpha}_s}\end{bmatrix},\end{equation}%
where
\begin{equation}
\begin{aligned}
	\mathbf{M}_{\theta_s\theta_s}&=2 \, \Re \left\{ \frac{\partial \bar{\mathbf{y}}_s^H}{\partial \theta_s} \mathbf{R}_n^{-1}\frac{\partial \bar{\mathbf{y}}_s}{\partial \theta_s} \right\}\notag\\
	&=\frac{2|\alpha_s|^2}{\sigma_n^2}\Re\left\{ \alpha_s\dot{\mathbf{g}}^H_s(\mathbf{p})\mathbf{F}\mathbf{S}\alpha^*_s\mathbf{S}^H\mathbf{F}^H\dot{\mathbf{g}}_s(\mathbf{p})\right\}\notag\\
	&=\frac{2|\alpha_s|^2L}{\sigma_n^2} \dot{\mathbf{g}}^H_s(\mathbf{p})\mathbf{F}\mathbf{F}^H\dot{\mathbf{g}}_s(\mathbf{p}),
\end{aligned}
\end{equation}
\begin{equation}
	\begin{aligned}
	\mathbf{M}_{\theta_s\boldsymbol{\alpha}_s}&=2 \, \Re \left\{ \frac{\partial \bar{\mathbf{y}}_s^H}{\partial \theta_s} \mathbf{R}_n^{-1} \frac{\partial \bar{\mathbf{y}}_s}{\partial \boldsymbol{\alpha}_s} \right\}\notag\\
	&=\frac{2}{\sigma_n^2}\Re\left\{\alpha_s\dot{\mathbf{g}}^H_s(\mathbf{p})\mathbf{F}\mathbf{S}\left( \boldsymbol{\epsilon}\otimes (\mathbf{S}^H\mathbf{F}^H\mathbf{g}_s(\mathbf{p}))\right)  \right\}\notag \\
	&=\frac{2L}{\sigma_n^2}\Re\left\{\alpha_s \dot{\mathbf{g}}^H_s(\mathbf{p})\mathbf{F}\mathbf{F}^H{\mathbf{g}}_s(\mathbf{p})\boldsymbol{\epsilon} \right\},\\
\end{aligned}
\end{equation}
\begin{equation}
	\begin{aligned}
	\mathbf{M}_{\boldsymbol{\alpha}_s\theta_s}&=2 \, \Re \left\{ \frac{\partial \bar{\mathbf{y}}_s^H}{\partial \boldsymbol{\alpha}_s} \mathbf{R}_n^{-1} \frac{\partial \bar{\mathbf{y}}_s}{\partial \theta_s} \right\}=\mathbf{M}^T_{\theta_s\boldsymbol{\alpha}_s},
\end{aligned}
\end{equation}
\begin{equation}
	\begin{aligned}
	\mathbf{M}_{\boldsymbol{\alpha}_s\boldsymbol{\alpha}_s}&=2 \, \Re \left\{ \frac{\partial \bar{\mathbf{y}}_s^H}{\partial \boldsymbol{\alpha}_s} \mathbf{R}_n^{-1} \frac{\partial \bar{\mathbf{y}}_s}{\partial \boldsymbol{\alpha}_s} \right\}\notag\\
	&=\frac{2}{\sigma_n^2}\Re\left\{\boldsymbol{\epsilon}^H\otimes (\mathbf{S}^H\mathbf{F}^H\mathbf{g}_s(\mathbf{p}))^H(\boldsymbol{\epsilon}\otimes (\mathbf{S}^H\mathbf{F}^H\mathbf{g}_s(\mathbf{p})))\right\}\notag\\
	&=\frac{2L}{\sigma_n^2}\Re\left\{\mathrm{Tr}\{{\mathbf{g}}^H_s(\mathbf{p})\mathbf{F}\mathbf{F}^H{\mathbf{g}}_s(\mathbf{p})\}\boldsymbol{\epsilon}^H  \boldsymbol{\epsilon} \right\}\notag\\
	&=\frac{2L}{\sigma_n^2}{\mathbf{g}}^H_s(\mathbf{p})\mathbf{F}\mathbf{F}^H{\mathbf{g}}_s(\mathbf{p})\mathbf{I}_2.
\end{aligned}
\end{equation}
Thus, the CRB for estimating $ \theta_s $ is given by
\begin{eqnarray}
	\begin{aligned}
		\mathrm{CRB}_{\theta_s} &=  \frac{1}{	\mathbf{M}_{\theta_s\theta_s}-\mathbf{M}_{\theta_s\boldsymbol{\alpha}_s}\mathbf{M}^{-1}_{\boldsymbol{\alpha}_s\boldsymbol{\alpha}_s}\mathbf{M}^T_{\theta_s\boldsymbol{\alpha}_s}}\\
		&=\frac{{\sigma_n^2}}{
			{2L|\alpha_s|^2}\left( 
			\dot{\mathbf{g}}^H_s(\mathbf{p})\mathbf{F}\mathbf{F}^H\dot{\mathbf{g}}_s(\mathbf{p}) 
			- \frac{ \left|   \mathbf{g}^H_s(\mathbf{p}) \mathbf{F} \mathbf{F}^H \dot{\mathbf{g}}_s(\mathbf{p})  \right|^2}
			{{\mathbf{g}}^H_s(\mathbf{p})\mathbf{F}\mathbf{F}^H{\mathbf{g}}_s(\mathbf{p}) }
			\right)
		}.
	\end{aligned}\label{crb}
\end{eqnarray}%

\end{document}